\documentclass[aps,showpacs,floatfix,superscriptaddress,prb,11pt]{revtex4-1}
\usepackage{amssymb,amsmath}
\usepackage{hyperref}
\usepackage{color}
\usepackage{graphicx}
\usepackage{ulem}

\def\beq{\begin{equation}}
\def\eeq{\end{equation}}

\usepackage{amsfonts}
\usepackage{epstopdf}
\usepackage{subcaption}

\begin{document}
\title{Anomalous Thouless energy and critical statistics on the metallic side of the many-body localization transition}
\author{Corentin L. Bertrand} 
\author{Antonio M. Garc\'ia-Garc\'ia}
\email{amg73@cam.ac.uk}
\affiliation{Cavendish Laboratory, University of Cambridge, JJ Thomson Av., Cambridge, CB3 0HE, UK}
\date{\today}
\begin{abstract}
We study a one-dimensional (1d) XXZ spin-chain in a random field on the metallic side of the many-body localization transition by level statistics. For a fixed interaction, and intermediate disorder below the many-body localization transition, we find that, asymptotically, the number variance grows faster than linear with a disorder dependent exponent. This is consistent with the existence of an anomlaous Thouless energy in the spectrum. In non-interacting disordered metals this is an energy scale related to the typical time for a particle to diffuse across the sample. 
In the interacting case it seems related to a more intricate anomalous diffusion process. This interpretation is not fully consistent with recent claims that, for intermediate disorder, level statistics are described by a plasma model with power-law decaying interactions whose number variance grows slower than linear. 
As disorder is further increased, still on the metallic side, the Thouless energy is gradually washed out. In the range of sizes we can explore, level statistics are scale invariant and approach Poisson statistics at the many-body localization transition. Slightly below the many-body localization transition, spectral correlations, well described by critical statistics, are quantitatively similar to those of a high dimensional, non-interacting, disordered conductor at the Anderson transition.   

\end{abstract}

\maketitle
\section{Introduction}
Spectral analysis is a powerful tool to probe the dynamics of non-interacting quantum disordered systems \cite{anderson1958,abou1973}. For instance Poisson statistics is a signature of a disordered insulator, a situation that occurs for any dimensionality for short-range hopping and sufficiently strong disorder \cite{frohlich1983}. By contrast, for disorder weak enough in more than two dimensions, the system is a disordered metal and the spectral correlations are given by Wigner-Dyson (WD) statistics \cite{efetov1983}. Deviations from WD statistics in disordered metals occur for eigenvalues separations larger than the Thouless energy \cite{altshuler1988}. Instead of the slow logarithmic increase of the number variance, typical of WD statistics,  a much faster power-law growth, with an exponent that only depends on the spatial dimensionality \cite{braun1995}, is observed beyond the Thouless energy. Its origin lies in the diffusive character of the motion for sufficiently short times.

As disorder increases, a metal-insulator transition takes place in more than two dimensions. Around the mobility edge the dynamics is characterized by universal critical exponents \cite{rodriguez2011,garcia2008}, anomalous diffusion \cite{ohtsuki1997}, a scale invariant dimensionless conductance \cite{abrahams1979} and multifractal eigenstates \cite{castellani1986,rodriguez2011}. Similar features are also found in other systems \cite{fishman1982,bogomolny1999,wang2009,garcia2006,garcia2008a,garcia2006a,garcia2007} where the potential is effectively quasi-random but ultimately deterministic. Spectral correlations are universal \cite{shklovskii1993} but different from WD or Poission statistics. Despite some initial controversy \cite{altshuler1988,shklovskii1993,kravtsov1994}, it is now clear that level statistics at the transition are characterized by the following features: a) scale invariance \cite{shklovskii1993} so there is no corrections due to the Thouless energy, b) level repulsion as in a metal, c) linear number variance, as for an insulator, but with a slope less than one that decreases with the space dimensionality \cite{schreiber1996,garcia2008,garcia2007a,markos2006}, d) the decay of the level spacing distribution for sufficiently large spectral separations is exponential, as for Poisson statistics, though with a different typical decay \cite{kravtsov1997}.

Generalized random matrix models, based on soft-confining potentials \cite{muttalib1993,nishigaki1999} or that are mapped onto the Calogero Sutherland model at finite temperature \cite{moshe1994,garcia2003}, have been successfully used to model the level statistics at the Anderson transition. In these models correlations between eigenvalues, usually called critical statistics \cite{kravtsov1997}, are suppressed exponentially with a typical decay that labels its universality class.  Also effective plasma models \cite{bogomolny1999}, where the correlations between eigenvalues are restricted to a finite number of eigenvalues, provide a qualitative description of the spectral correlations at criticality. For spectral interactions restricted only to nearest neighbours level statistics are termed semi-Poisson though this name is sometimes used to refer to the general case where the nearest k-eigenvalues are correlated. %Plasma models \cite{kravtsov1995} with %power-law decay correlations predict a slower than linear growth of the number variance. Therefore they do not describe %well the leading behaviour of the spectral correlations at the transition.
The upshot of this discussion is that level statistics, that requires much less computational effort than observables involving eigenfunctions, provide a rather complete description of the relevant physics of these systems.

A natural question to ask is whether level statistics are also helpful to characterize the dynamics of a disordered system in the presence of interactions. Indeed the description of the interplay between disorder and interactions \cite{fleishman1980,shepelyansky1994}, loosely referred to as many-body localization (MBL) \cite{basko2006}, has attracted enormous interest in recent years. We summarize below the main properties of this novel state of quantum matter. In Refs. \cite{basko2006,basko2006a} it was found, based on the approximate analytical treatment, that for sufficiently strong disorder, Anderson localization in the non-interacting problem is robust to weak interactions. A direct consequence is that, neglecting phonons, the DC conductivity is strictly zero for sufficiently strong disorder and low temperatures. Rigorous mathematical results have confirmed this prediction in some limiting cases. For a vanishing density, localization persists \cite{aizenman2007} if weak interactions are turned on. In the limit of mean-field interactions, where the Hamiltonian is just the non-linear Schroedinger equation, it was demonstrated rigorously \cite{wang2008} that  weak interactions in an otherwise Anderson insulator induce at most a very-slow logarithmic like diffusion. A proof of MBL in a one dimensional spin chain has been recently reported \cite{imbrie2016,imbrie2016a}.
Numerical simulations in small volumes suggest that the insulating side of the many-body localization transition is characterized by a logarithmic 
\cite{znidaric2008,bardarson2012}, instead of linear, growth of the entanglement entropy, zero dc conductivity and a faster than linear growth of the ac conductivity in the low frequency limit \cite{gopalakrishnan2015}. 

On the metallic side, but close to the transition, it has been identified a Griffith-like phase \cite{bar2014,bar2015,luitz2015,agarwal2015,luitz2016,znidaric2016,vosk2015,gopalakrishnan2016} characterized by slow sub-diffusion and also a sub-linear power-law growth of the entanglement entropy.
Not much is known for sure about the physics around the MBL transition. Numerical calculations of the critical exponents \cite{luitz2015} suggest a violation of the Harris criterion however this result has been recently challenged \cite{chandran2015}. In any case, even if the violation does occur, this is not necessarily incorrect in this context \cite{monthus2015}. Due to similarities with the physics of a single particle on a Bethe lattice \cite{aizenman2009,biroli2012,deluca2014,basko2006}, it is plausible that level statistics around the transition are close to Poisson statistics as for an insulator. 

%This is also consistent with numerical simulations \cite{luitz2015} though the volumes are too small to be sure %that this is the definitive answer. 
Regarding level statistics, a detailed analysis of the 
spectral correlations of a one-dimensional XXZ chain with a random field in the strong disorder limit revealed \cite{kudo2004} a transition from Poisson to WD statistics as the system crosses the MBL transition by tuning interactions. In a more recent paper \cite{serbyn2016}, the metallic side of the many-body localization transition in a similar chain was also investigated by level statistics. It was found for intermediate disorder, still far from the transition, level statistics are well described by an effective eigenvalue plasma model with power-law interactions \cite{kravtsov1995} leading to a slower than linear growth of the number variance. Close to the MBL transition, but still on the metallic side, it was reported that spectral correlations, described by semi-Poisson statistics, are similar to those of a disordered conductor at the Anderson transition.

Here we revisit the study of level statistics in the critical region around the MBL transition and for intermediate disorder deep in the metallic phase. As in the non-interacting limit, we aim to characterize the quantum dynamics by a technically simpler and basis invariant spectral analysis. 

For sufficiently weak disorder we identify the Thouless energy \cite{braun1995}, typical of a disordered metal in a finite size box, in the spectrum. However, unlike the non-interacting case, in this case it is related to sub-diffusive and disorder dependent dynamics. 
For eigenvalue separations smaller than the Thouless energy, the number variance grows logarithmically as in WD statistics. For larger separations, the growth is power-law and faster than linear with an exponent that decreases with disorder. This suggests that the dynamics not too close to the transition is controlled by a process of anomalous diffusion. The level spacing distribution show size-dependent deviations from WD statistics also consistent with the existence of an anomalous Thouless energy. We note that in Ref.\cite{serbyn2016} the growth of the number variance, was found to be slower than linear. This difference is important for the physical interpretation of the results. A faster than linear power-law growth of the number variance is a signature of the Thouless energy, namely, a feature of a non-critical metal. By contrast a slower than linear growth, at least in the non-interacting case, is a feature associated to criticality that can only occur in system with a scale-invariant spectrum close to the Anderson transition. As we will explain in detail, our discrepancy with Ref.\cite{serbyn2016} is rooted in the different spectral windows employed to compute the number variance. While in Ref.\cite{serbyn2016} only spectral windows containing up to $20$ eigenvalues were considered, here we study energy intervals with up to $300$ eigenvalues. We believe that this is necessary for a more accurate account of power-law growth.

% A plasma model \cite{jalabert1993} with logarithmic interactions for short range correlations and power-law %interactions for long-range interaction captures well the spectral correlations in this region. 
As the MBL transition is approached from the metallic size, the Thouless energy becomes harder to observe. Level statistics undergo a smooth crossover towards Poisson statistics. Slightly below the transition, spectral correlations, scale-invariant and well described by critical-statistics, are strikingly similar to those of a high-dimensional non-interacting disordered system at the Anderson transition.  
Unlike Ref.\cite{serbyn2016} we do not observe any signature of semi-Poisson statistics, as defined in Ref.\cite{bogomolny1999}, close to the MBL transition. However we agree with Ref.\cite{serbyn2016} that in this region level statistics are critical as in a non-interacting disordered system at the Anderson transition.

We start by introducing the model and giving some details of the numerical analysis.
\section{The model and the numerical analysis}
We study the one-dimensional (1d) XXZ Heisenberg model in a random magnetic field:
\begin{equation}
\label{eq:hamiltonian}
H = \sum_{i=0}^{L-1} {\hat{S}}_i \cdot  {\hat{S}}_{i+1} + w_i \hat{S}_i^z
\end{equation}
where $\hat{S}^{x,y,z} = \frac{1}{2}\hat{\sigma}^{x,y,z}$, $\hat{\sigma}$ denotes the Pauli matrices and $w_i$ is a random magnetic field with a uniform distribution $\left[-h, h\right]$.
We consider $L=12, 14, ..., 18$ spins. The dimension of the Hilbert space is therefore $2^L$. We employ periodic boundary conditions in order to minimize finite size effects.\\

We note it is important that the symmetry of the eigenstates considered is the same. Following previous literature, we focus on the subset of eigenvalues associated to eigenstates of the operator $\hat{S}^z=\sum_i \hat{S}_i^z$. To keep a maximum number of eigenvalues, only the channel $S_z = 0$ is considered, where $S_z$ is the eigenvalue of $\hat{S}_z$. 
%Note that $L$ and $2S$ cannot have different parities.\\

Eigenvalues were computed with a routine of the library Eigen\footnote{Eigen - a C++ template library for linear algebra: matrices, vectors, numerical solvers, and related algorithms. http://eigen.tuxfamily.org}. Eigen was chosen because its computation time for diagonalisation is shorter than LAPACK for equal accuracy. \\
The whole spectrum was calculated as Eigen does not implement partial diagonalisation.
% Eigen uses a QR {\bf what is QR?} algorithm taking advantage of the hermicity of the Hamiltonian. According to Eigen documentation, its complexity is $9n^3$, where $n$ is the %order of the Hamiltonian. When restricted to the channel $S$, $n$ is the binomial coefficient with parameters $L$ and $(L-2S)/2$. 
The maximum size $L=18$ we could explore numerically was mostly dictated by the $128$GB RAM memory available. Table~\ref{tab:nb_eigenvalues} provides detailed information on the number of disorder realisations and the total number of eigenvalues for each value of the disorder $h$.\\
	
	\begin{table}%
		\centering
		\begin{tabular}{|c|c|c|}
			\hline
			$L$ & Disorder Realisations & Number of eigenvalues \\
			\hline
			12 & 10000 & 9240000 \\
			14 & 1000 & 3432000 \\
			16 ($h\geq 2$) & 212 & 2728440 \\
			16 ($h = 2$) & 100 & 1287000 \\ 
			18 ($ 2 \leq h \leq 4$) & 65 & 3160300 \\
			18 ($h = 0.75,1.25,1.75$) & 24 & 1166880 \\
		    18 ($h = 0.5,1,1.5$) & 18 & 875160 \\
			\hline
		\end{tabular}
		\caption{Number of disorder realisations and eigenvalues calculated for each system size $L$ and disorder $h$ considered.}
		\label{tab:nb_eigenvalues}
	\end{table}
For disorder below the critical value $h_c$, extended and localized states must not be mixed in the analysis of the spectrum. Moreover the mobility edge is smeared out due to finite size effects. We check that for all the considered disorder strength and sizes, it is safe to take one eighth of the spectrum around the center, namely, $12.5$\% of the full spectrum.

%the part of the spectrum in which the level statistics analysis is carried out only contains localized, extended %or critical eigenstates.  calculated the number variance for the centre 12.5\% and the centre 25\% of the spectrum 5in the vicinity of the %transition. There was no significant difference for $N < 500$ as seen in Fig.~\ref{fig:comparison_spectrum_part}. Consequently, the centre 12.5\% was considered as ``safe'', and this part of the spectrum was taken for all calculations.\\
\begin{figure}%
	\centering
	\resizebox{0.95\textwidth}{!}{\includegraphics{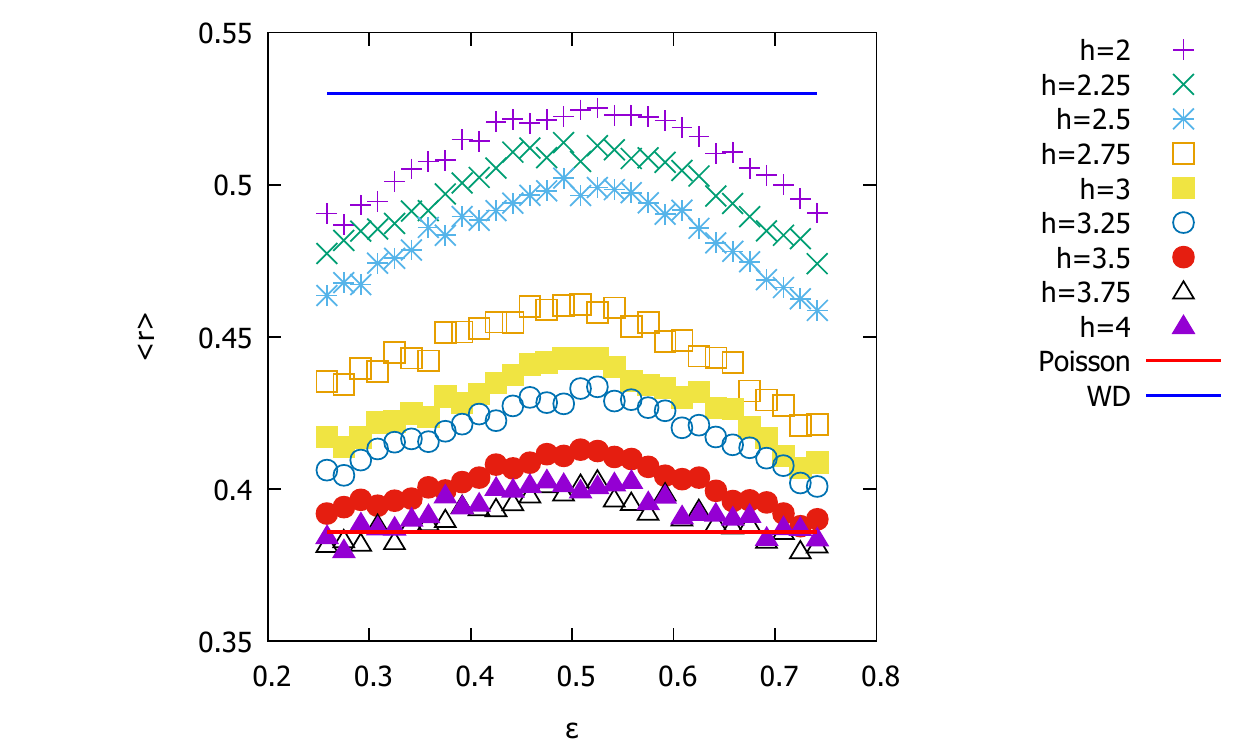}}
	\caption{Average adjacent gap ratio $\langle r \rangle$ for the centred half of spectrum for $L=18$ and different values of disorder where $\epsilon = \frac{E - E_{\rm min}}{E_{\rm max} - E_{\rm min}}$ where $E_{\rm min}$ and $E_{\rm max}$ are the minimum and maximum eigenvalues in the spectrum. The centre of the spectrum is close to WD statistics for $h \leq 2.5$, while for $h \geq 3.5$ is closer to the Poisson statistics prediction. The MBL transition $h = h_c$ at the center of the spectrum occurs between these two values.}%
	\label{fig:agr}%
\end{figure}%

\subsection{Unfolding}
\label{sec:unfolding}
The averaged spectral density is highly non-universal and, in general, it does not give direct information on the quantum dynamics.
 It is therefore common that in level statistics studies the average spectral density is extracted from the numerical spectra. This procedure, termed unfolding \cite{guhr1998}, 
consists in a local rescaling of the spectrum so that the local density of states is one (details in appendix~\ref{ap:unfolding}). More specifically, the numerical spectral density, generically a fluctuating quantity, is fitted by a smooth function. This smooth spectral density is then used to rescale the spectrum. 
In that way it is also possible to compare eigenvalues from different parts of the spectrum as distances in the spectrum are measured in units of the local mean level spacing.

Unfolding is a rather delicate task as it requires to carry out a careful separation between smooth and fluctuating parts of the spectral density. 
To remove unfolding artefacts we employ two local and two global fitting methods (see appendix~\ref{ap:unfolding} for a detailed comparison). 
%Results are shown in the appendix Fig.~\ref{fig:comparison_unfolding}. 
In general, local fittings are more accurate for neighbouring eigenvalues but tend to destroy long-range correlations that control long range spectral observable such as the number variance. By contrast, global fittings are less accurate in extracting the smooth part of the density but conserve long-range correlations. After a careful comparison between the different methods we have opted for an average cubic global fitting for all the results presented in the paper. It agrees with a local fitting for short-range spectral correlations and, unlike the global simple cubic unfolding, the average spectral density is still very close to that predicted by a local unfolding. Moreover we have checked that this unfolding reproduces quantitatively previous results \cite{braun1995} on long-range spectral correlations in a non-interacting three dimensional tight-binding disordered system. 

\subsection{Critical disorder $h_c$ and critical exponent $\nu$}
As a first step in the study of spectral correlations, we employ the adjacent gap ratio (\ref{eq:agr}) \cite{luitz2015,oganesyan2007}, 

\begin{equation}
	\begin{split}
		&r_i = \frac{\min(\delta_i, \delta_{i+1})}{\max(\delta_i, \delta_{i+1})} \\
		&\delta_i = E_i - E_{i-1},
	\end{split}
	\label{eq:agr}
\end{equation}
where it is assumed that the spectrum is ordered $E_{i-1} < E_i < E_{i+1}$, in order to determine the critical disorder $h_c$ at which the MBL transition occurs. 
 
 \begin{figure}%
 	\centering
 	\resizebox{0.95\textwidth}{!}{\includegraphics{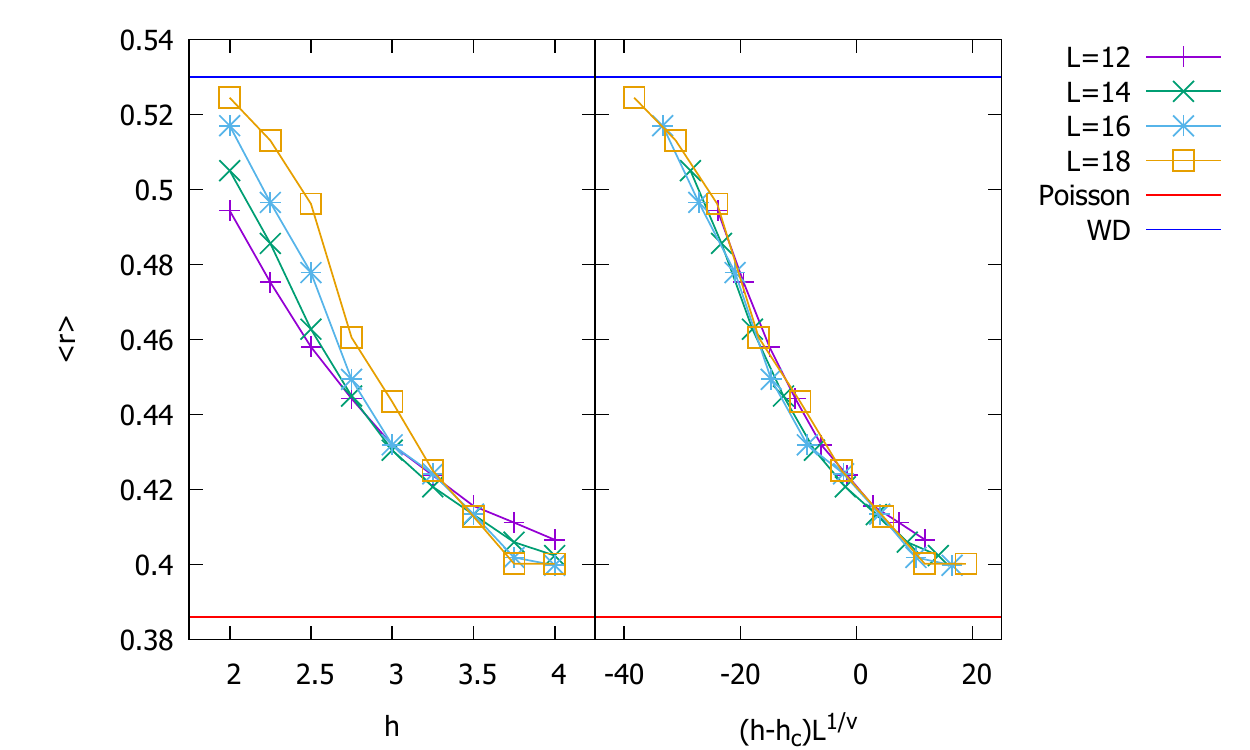}}
 	\caption{Finite-size scaling analysis of the adjacent gap ratio $\langle r \rangle$ after ensemble average. Left: Disorder dependence of the averaged adjacent gap ratio $\langle r \rangle$ (\ref{eq:agr}) for different sizes. The crossing point $h = h_c$ corresponds to the mobility edge of the MBL transition. Right panel: Rescaled adjacent gap ratio so that all curves for different sizes collapse in a single curve. The best fit corresponds to $h_c = 3.35 \pm 0.05$ and $\nu = 0.86 \pm  0.13$ where $\nu$ is the critical exponent that controls deviations from spectral scale-invariance and also the divergence of the localization length (see appendix~\ref{ap:scaling_analysis} for more details).}%
 	\label{fig:scaling}%
 \end{figure}
 
The average adjacent gap ratio for a Poisson distribution is $\left\langle r \right\rangle_P = 2\ln(2) - 1 \approx 0.386$. For the WD distribution, corresponding to a disordered metal, it is $\left\langle r \right\rangle_W \approx 0.530$. We note that throughout the paper we define {\it disordered metal}  not as a state of matter with finite conductivity but rather as a state of matter in which some degree of level repulsion persists. 

 At the Anderson or MBL transition $h \equiv h_c$, it lies in between these two values even in the thermodynamic limit \cite{oganesyan2007}. This observable has two advantages: it is local, so it is less affected by size effects and it is a ratio between local quantities, so it is independent from the unfolding procedure. In Fig.~\ref{fig:agr} we depict results for the adjacent gap ratio for the centred half of the spectrum and different disorder strength. As was expected, for sufficiently weak disorder $h \leq 2.5$, and close to center, the adjacent gap ratio is close to the WD prediction while in the strong disorder limit, $h \geq 3.5$ it is already very close to the Poisson value typical of an insulator. The MBL transition around the center of the band must therefore occur for intermediate values of disorder $2.25 < h_c < 4.0$.  

We note that, because of finite size effects, the mobility edge at $h_c$ is not sharp, there exists a size-dependent region around it with critical properties.
In order to estimate $h_c$ and $\nu$ (see more details in appendix~\ref{ap:scaling_analysis}) it is therefore necessary to carry out  (see Fig.~\ref{fig:scaling}) a finite-size scaling analysis of the adjacent gap ratio.

At the MBL transition the localization length diverges and the spectral correlations are scale invariant, which is used to determine $h_c$ (crossing point in the left panel). Sufficiently close to the MBL transition level statistics are controlled by scaling laws. It is therefore expected that, after a proper rescaling, the adjacent gap ratio for different sizes will collapse in a single curve. The scaling for which this occurs allows us to estimate the critical exponent $\nu$ (right panel) that controls both deviations from spectral scale-invariance  and the divergence of the localization length at the transition.   
The results of this analysis are $h_c \approx 3.35 \pm 0.05$ and $\nu \approx 0.86\pm 0.13$. The error bars were estimated by using a random fitting range (see details in appendix~\ref{ap:scaling_analysis}) and a random starting value for the minimum finder. The error in $h_c$ and $\nu$ is the standard deviation after averaging over $10000$ realisations. While the resulting error estimation for $\nu$ seems reasonable the one for $h_c$ is not realistic given other systematic uncertainties. These results are in line with those of Ref.\cite{luitz2015}  $h_c \approx 3.6$ and $\nu = 0.8 \pm 0.3$ that considered larger sizes and had better statistics. The slightly smaller $h_c$ in our case is likely due to the fact that we are considering a larger spectral window ($12\%$) to compute it. In \cite{luitz2015} it was explicitly found that $h_c$ decreases as one moves from the center of the spectrum. We note  $\nu \approx 1$ is also the critical exponent that controls the divergence of the localization length of a non-interacting particle in a disordered Cayley tree at the Anderson transition. Interestingly there are striking similarities \cite{basko2006} between these two problems. 

\section{Results}
We have now all the ingredients to compute the spectral correlations of the Hamiltonian (\ref{eq:hamiltonian}).
More specifically we study the level spacing distribution (\ref{eq:spacing_distribution}), a short range spectral correlator, and the number variance (\ref{eq:nv}), a long-range spectral correlator, that provide valuable insights on the quantum dynamics in the long and short time limits respectively.  We start with the latter as this is the one more suited to investigate the existence of the Thouless energy in the system which is one of the main goals of the paper. 

\begin{figure}%
	\centering
	\resizebox{0.95\textwidth}{!}{
		\resizebox{0.9\textwidth}{!}{\includegraphics{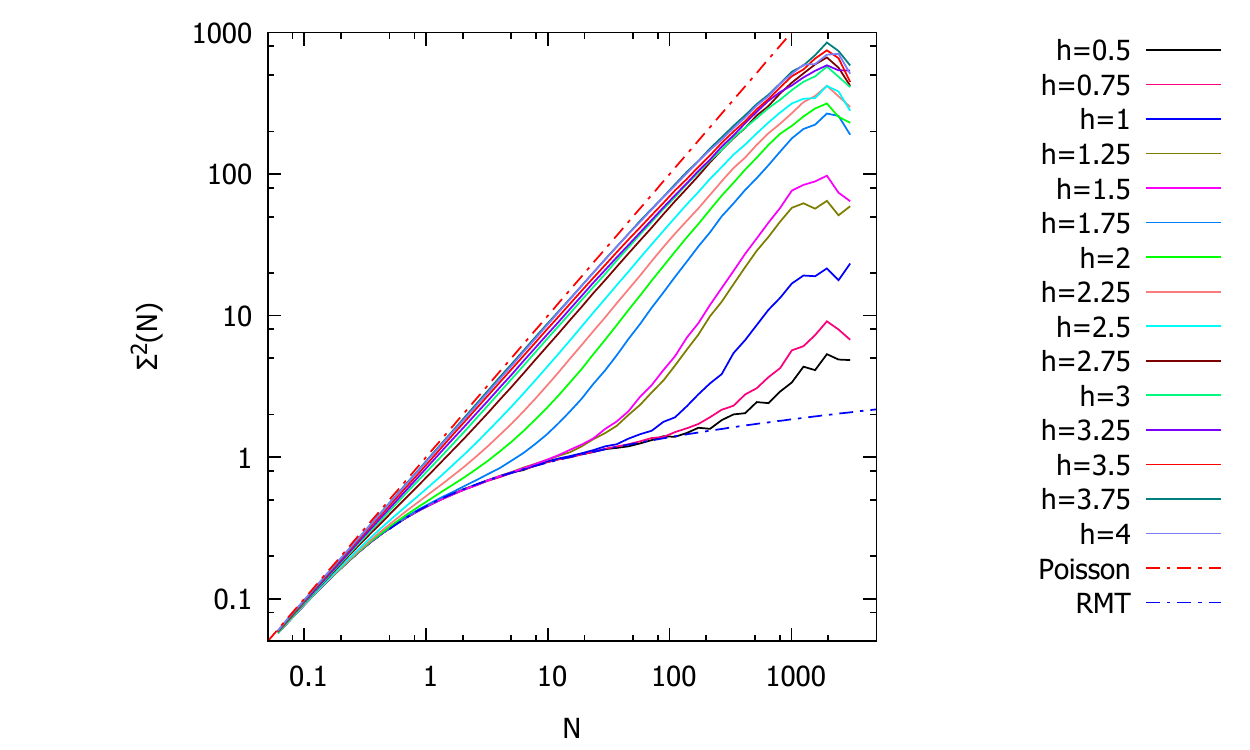}}
		%\resizebox{0.3\textwidth}{!}{\input{variance_exponents}}
	}
	\caption{Number variance $\Sigma^2(N)$ (\ref{eq:nv}) for a fixed size $L = 18$ and different disorder strength $h$ across the MBL transition. For small $N$ and weak disorder, we clearly observe a logarithmic growth typical of a disordered metal. For larger $N$, the observed faster than linear power-law growth is a signature of the Thouless energy in the system. The saturation observed for even larger $N$ is a consequence of the finite number of levels used to compute $\Sigma^2(N)$. For stronger disorder ($h > 2.5$) the number variance seem to approach Poisson statistics for all $N$, as in an Anderson insulator, even on the metallic side of the transition $h < h_c \approx 3.4$. In section \ref{compa} we compare deviations form Poisson with the predictions of critical statistics.}%
	\label{fig:variance}%
\end{figure}%

\begin{figure}%
	\centering
	\resizebox{0.9\textwidth}{!}{
		\resizebox{0.9\textwidth}{!}{\includegraphics{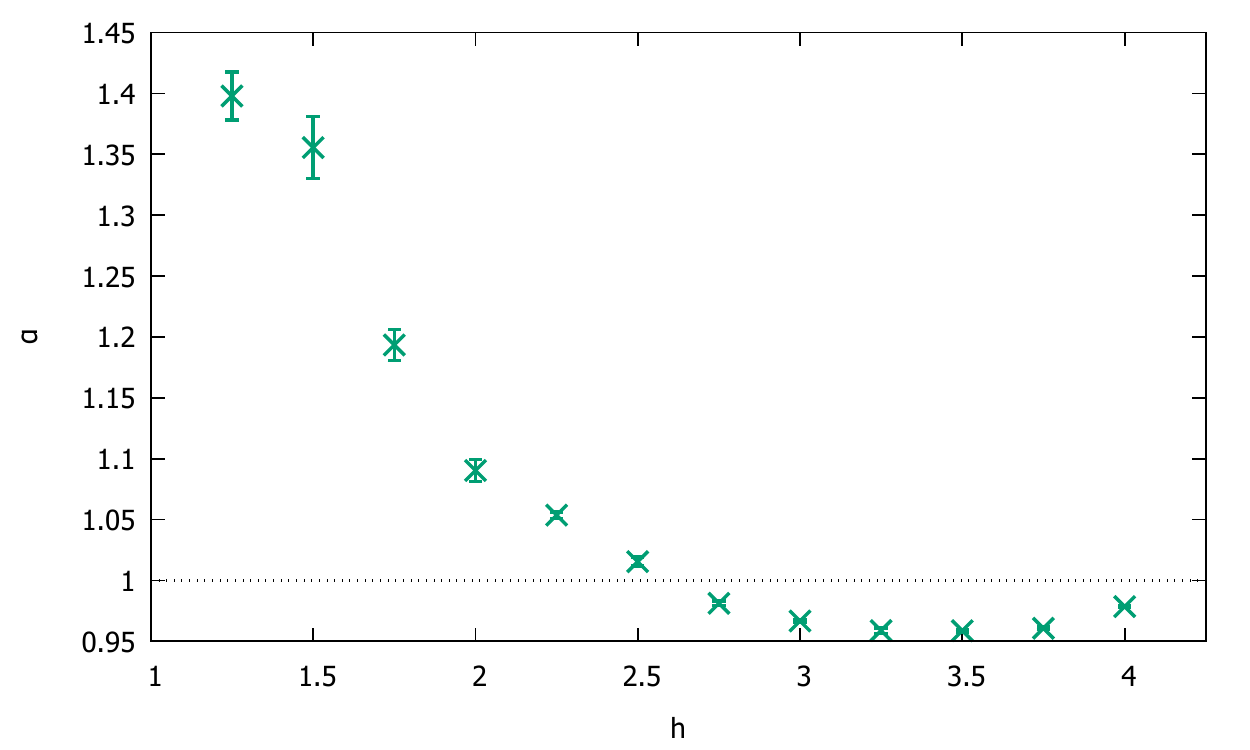}}
		%\resizebox{0.3\textwidth}{!}{\input{variance_exponents}}
	}
	\caption{Power-law exponent $\alpha$ that controls the asymptotic growth of the number variance $\Sigma^2(N) \sim N^\alpha$ across the transition. We used a fitting function $a+bN^\alpha$ with $a,b,\alpha$ fitting parameters. For intermediate disorder ($h \leq 2.5$) the faster than linear ($\alpha > 1$) growth signals the existence of the Thouless energy. This does not agree with the results of \cite{serbyn2016}  (see Fig. \ref{fig:variance2} for more details). As the MBL is approached $h_c \approx 3.35$ the growth becomes linear with small deviations likely due to finite size effects.  These results, especially the faster than linear growth for ($h \leq 2.5$) are robust to changes in the fitting interval $[N_i,N_f]$ provided that roughly $N_f -N_i \geq 100$ and $N_i \gg 1$, $N_f \leq 300$.}
	\label{fig:exponent}%
\end{figure}%

\begin{figure}%
	\centering
	\resizebox{0.99\textwidth}{!}{\includegraphics{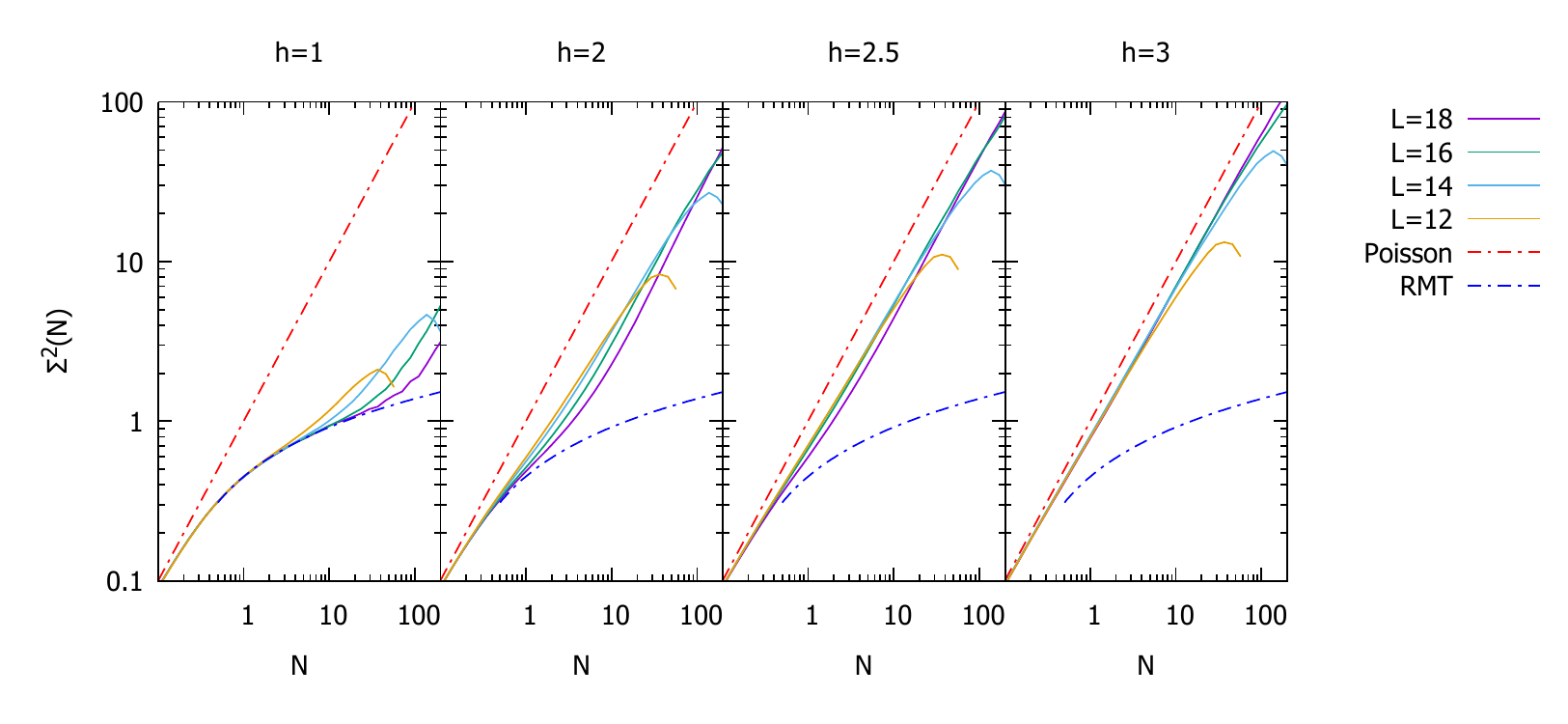}}
	\caption{Size dependence of $\Sigma^2(N)$  (\ref{eq:nv}) for different disorder strength $h$. Deviations from logarithmic growth start to occur for larger $N$ as size increases. This is another indication that the power-law growth of $\Sigma^2(N)$ for larger $N$ is a signature of the existence of the Thouless energy in this system. For $h > 2.5$, the number variance is almost linear $\Sigma^2(N) \approx \chi N$ and scale-invariant. It seems to be close to Poisson statistics as $h \to h_c$. No Thouless energy is observed in this region.}% We did not find any evidence of the semi-Poisson statistics $\chi = 1/2$ around the transition predicted in \cite{serbyn2016}}%
	\label{fig:variance_scalefree}%
\end{figure}%

	\subsection{Long range spectral correlations: the number variance}	
	We start by investigating long-range spectral correlations that provide information on shorter time dynamics which is more sensitive to the existence of the Thouless energy in the system. 
	For that purpose we employ the number variance defined as the variance of the number of levels $N(\epsilon)$ lying in a band of energy $\epsilon$ (in units of the mean level spacing):
	\begin{equation}
		\Sigma^2(\epsilon) = \left\langle N^2(\epsilon)\right\rangle - \left\langle N(\epsilon)\right\rangle^2.
		\label{eq:nv}
	\end{equation}
	Since in the unfolded spectrum the average spectral density is $1$, $\left\langle N(\epsilon)\right\rangle = \epsilon$, the energy parameter $\epsilon$ can be replaced by the average number of levels $\left\langle N\right\rangle$, denoted by $N$ for simplicity.\\
	
	For a Poisson distribution typical of an insulator, different parts of the spectrum are not correlated, so the number variance is linear with slope one $\Sigma^2(N) = N$. 
	By contrast, in a disordered metal or in a random matrix \cite{mehta2004} level repulsion causes, for $N \gg 1$, a slow logarithmic increases of the number variance:
	\begin{equation}
		\Sigma^2(N) \approx \frac{2}{\pi^2}\log(N).
		\label{eq:nvwd}
	\end{equation}
	This slow growth of the number variance, in comparison with that for Poisson statistics, illustrates another spectral signature of disordered metals: spectral rigidity.\\
	In 
	Fig.~\ref{fig:variance} we depict results for the number variance for $L = 18$ as a function of the disorder strength $h$. For sufficiently weak disorder $h \ll h_c$, and small $N$, we clearly observe the logarithmic growth expected in a disordered metal. However, as the eigenvalue separation $N$ increases further, the number variance undergoes a slow crossover to a much faster power-law growth $\Sigma^2(N) \sim N^\alpha$ $\alpha(h) \geq 1$ (see Fig.~\ref{fig:exponent}). This is a clear signature of the Thouless energy \cite{braun1995} in the system. 
	
	As was mentioned in the introduction, the Thouless energy in the non-interacting limit is a energy scale related to the time that the particle takes to cross the sample. Assuming normal diffusion, still in the non interacting limit, the growth of the number variance is expected to grow as $N^{d/2}$ independently on disorder, where $d>2$ is the space dimensionality. These arguments cannot easily be carried over to the interaction case though, on physical grounds, we also expect that anomalous diffusion, that characterizes the disorder dependent Griffith phase, will lead to an anomalous growth of the number variance for sufficiently large $N$, corresponding to small times of the quantum evolution. It has been argued \cite{basko2006,deluca2014} that the MBL problem share similarities with a single particle in a $d \to \infty$ lattice. If this analogy is applicable here, it would correspond to an exponential growth of the number variance for energies larger than the Thouless energy.

	Indeed, see Fig.~\ref{fig:exponent}, we observe that, for not too weak disorder, the growth of the number variance is faster than linear, but seem to be slower than exponential. We note that it is not possible to completely rule out an exponential growth. It is well known \cite{braun1995} that, in the non-interacting problem, the asymptotic form of the growth $\sim N^{d/2}$ is only achieved for energies much larger than the Thouless energy which cannot be reached in the present numerical simulation. Moreover size effects induced by both the finite lattice size and the finite number of eigenvalues used to compute the number variance will suppress the growth. The leading correction due to the latter, expected to be $\sim  - N^2/n$ where $n$ is the number of eigenvalues in a single realization of disorder, is expected to become relevant for $N >200 ~ (L=18)$.
	
	The existence of the Thouless energy is further confirmed in Fig.~\ref{fig:variance_scalefree} that depicts the number variance for a fixed disorder and different sizes. The threshold for the observation of the power-law on the metallic side increases with system size while the exponent does not seem to depend on $L$.  
	
	As disorder approaches $h_c$, the Thouless energy gradually disappears. Even for small $N$ we do not observe WD statistics. On the metallic side, but close to $h_c$, the number variance is linear $\Sigma^2(N) \approx \chi N$ with a slope $\chi \lesssim 1$ that increases as $h \to h_c$.

	%For weak disorder $h \leq 1.5$ the growth seems to be slower than linear, though this is %probably a numerical artefact. Deviations from the logarithmic growth only occur at large %$N \gg 100$, a region which is especially sensitive to unfolding artefacts and finite-size %effects. Also, less data was calculated in this region {\bf remove this region if we are %not sure?}.

	These results contrast with those of Ref. \cite{serbyn2016} where it was found that for intermediate disorder level statistics are described by a plasma-model, with power-law interactions, leading a sublinear growth of the number variance for intermediate disorder. 
\subsection{Short range spectral correlations: the level spacing distribution}
The level spacing distribution $P(s)$ provides useful information about short-range spectral correlations related with the system evolution for long times of the order of the Heisenberg time, a time scale related to the inverse of the mean level spacing $\Delta$. More specifically it is the probability to find two eigenvalues separated at a distance $s$ in units of $\Delta$ with no other eigenvalues in between:
\begin{equation}
P(s) = \sum_i \left\langle \delta(s-\epsilon_i+\epsilon_{i+1})\right\rangle \ \ \  \epsilon_i = E_i/\Delta,
\label{eq:spacing_distribution}
\end{equation}

In an insulator it is given by Poisson statistics:
\begin{equation}
P_P(s) = e^{-s}.
\label{eq:poisson}
\end{equation}
By contrast in a disordered metal for $L \to \infty$ it is given by  
the WD spacing distribution \cite{mehta2004} which is well approximated by, 
\begin{equation}
P_W(s) \approx \frac{\pi}{2}s\exp(-\frac{\pi}{4} s^2).
\label{eq:wd}
\end{equation}
Unlike Poisson statistics, $P_W(s) \sim s$ for $s\rightarrow 0$. This level repulsion is characteristic of extended states.

\begin{figure}%
	\centering
	\resizebox{0.95\textwidth}{!}{\includegraphics{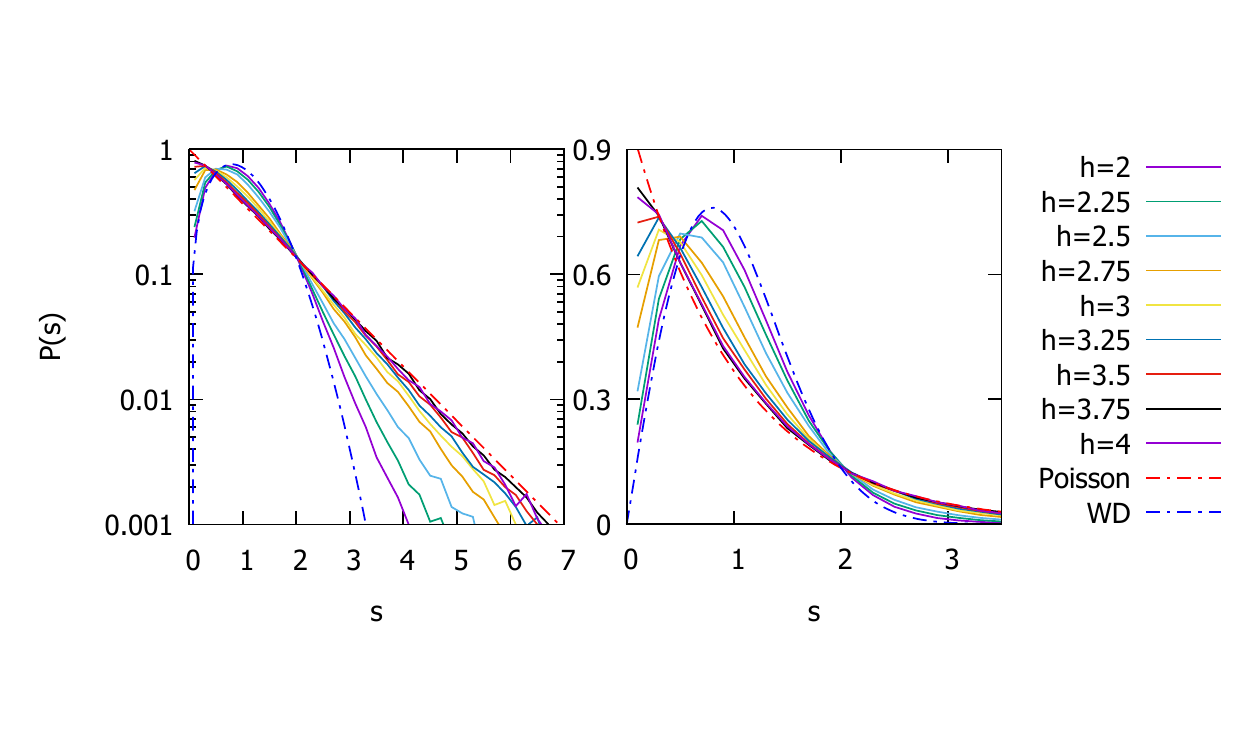}}
	\caption{Level spacing distribution $P(s)$ (\ref{eq:spacing_distribution}) for $L=18$ as a function of the disorder. A crossover from WD to Poisson statistics is clearly observed as disorder increases.}%
	\label{fig:spacings}%
\end{figure}

In Fig.~\ref{fig:spacings} we depict results of $P(s)$ as a function of disorder $h$ for $L=18$, the largest size that we can reach numerically. As was expected, it is close to Poisson statistics for sufficiently large $h > h_c$ while for small $h < h_c$ it is close to WD statistics. In the critical region $h \approx h_c$ it is closer to Poisson statistics but level repulsion is still observed for small energy separations $s \ll 1$. Moreover small deviations are also observed for $s \gg 1$. Much larger sizes would be necessary to determine whether these deviations are a finite size effect or a genuine feature of the scale-invariant $P(s)$ that describes the MBL transition. Taking into account the small size that is possible to explore numerically, and the proximity of $P(s)$ to Poisson statistics as $h \to h_c$, we tend to believe that this is just a size effect. Even if this is the case it would be interesting to characterize it more quantitatively. We will do that in section \ref{compa}. 

\begin{figure}%
	\centering
	\resizebox{0.95\textwidth}{!}{\includegraphics{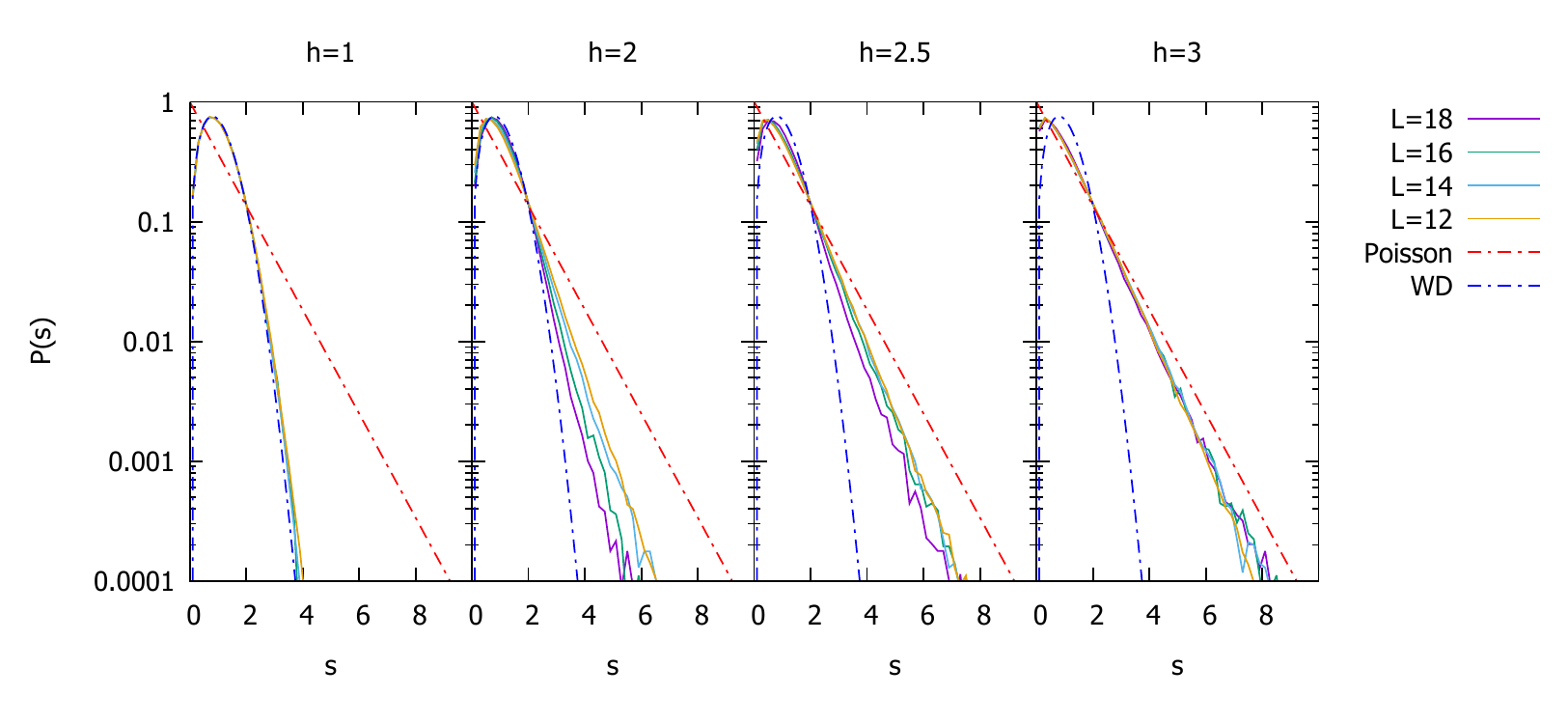}}
	\caption{Size dependence of $P(s)$ for different disorder strength $h$ in the metallic phase. Well in the metallic phase the scaling is consistent with that of a disordered metal with a Thouless energy in the spectrum. As the system approaches the MBL transition $h_c \approx 3.4$, $P(s)$ becomes scale invariant. $P(s)$ is closer to Poisson statistics but the decay is slightly faster.}%
	\label{fig:spacing_scalefree}%
\end{figure}%
At the moment we explore the limits of this (pseudo)-criticality by studying the size dependence of $P(s)$ for a given disorder $h$. The results, depicted in Fig.~\ref{fig:spacing_scalefree}, are fully consistent with the previous finite size scaling analysis of the adjacent gap ratio. Deep in the metallic region, $h < 2.5$, $P(s)$ is size-dependent and becomes closer to WD for larger sizes, namely, it is more metallic. The point from which deviations from WD start to appear increases with system size. Although the statistics is not good enough to make definite claims, this is an early signature of the Thouless energy in this system. By contrast, in the insulating region $h > h_c$ (not shown) $P(s)$ is well described by Poisson statistics for all the sizes considered.

The level spacing distribution $P(s)$ in the metallic side, but close to the transition $2.5 \leq h < 3.25$, seems to be scale invariant. This is a signature of criticality, namely, the localization length is already larger than the system size and the system behaves as if it is already at the MBL transition. Although it is close to Poisson we still clearly observe level repulsion for $s \ll 1$ and deviations from Poisson statistics for larger energy separations. For $h = 2.5$ it seems that the decay is slightly faster than exponential though we refrain from making a more quantitative statement because the range of spectral distances $s < 9$ available is too small.

\begin{figure}%
		\centering
		\resizebox{0.9\textwidth}{!}{
			\resizebox{0.9\textwidth}{!}{\includegraphics{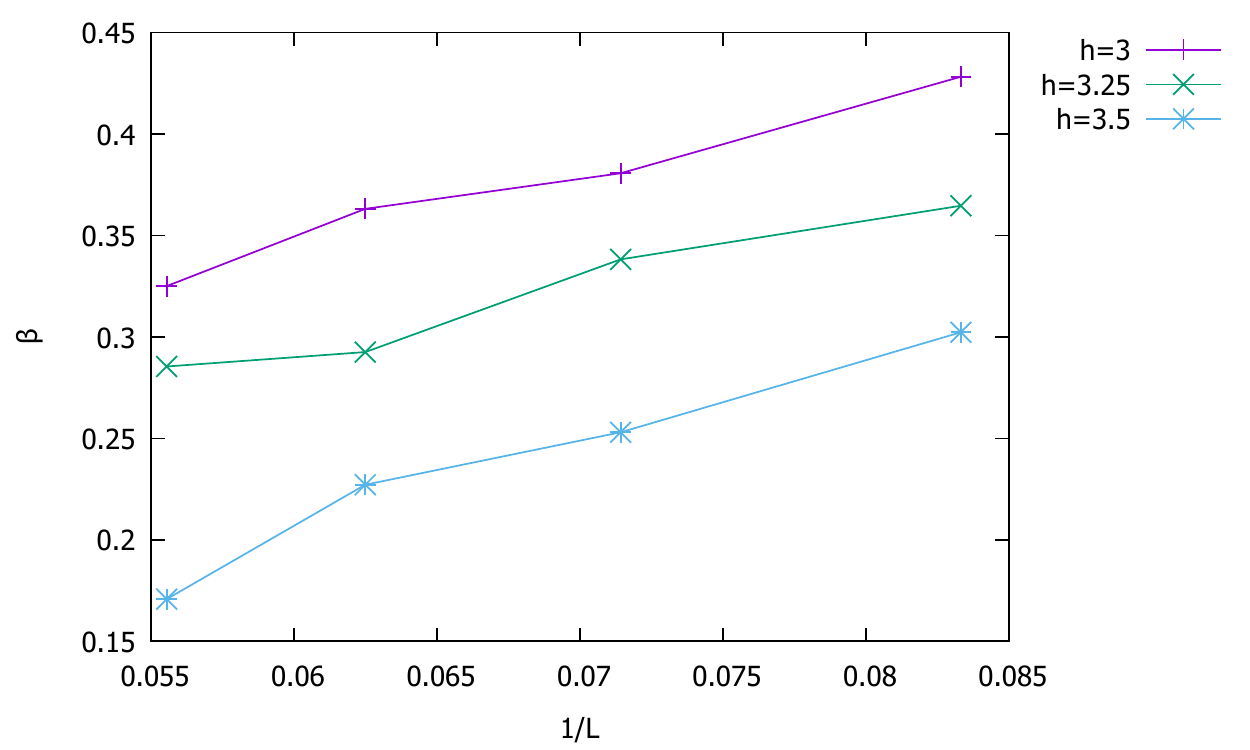}}
			%\resizebox{0.3\textwidth}{!}{\input{variance_exponents}}
		}
		\caption{Power-law exponent $\beta$, that controls the strength of level repulsion, $P(s)\sim s^\beta$ for $s \ll1 $, around the MBL transition as a function of the system size $L$. The exponent $\beta$ decreases as disorder $h$ and $L$ increasis. For $L = 18$ and $h \sim H\sim h_c$ is already much smaller than one. This is a strong indication that at the MBL transition level statistics are close to Poisson statistics $\beta \approx 0$. Larger volumes are needed to fully confirm this prediction.}
		\label{fig:psbeta}%
	\end{figure}

A similar non-exponential decay was observed in early numerical studies \cite{varga1995,evangelou1994,guhr1998} of $P(s)$ close to the Anderson transition in the non interacting limit. In that case it turned out that this was only a numerical artefact due to the small size of the system and the limited range of spectral distances ($s$) that could be explored numerically. Although further studies are required to clarify this issue, it is possible this might also be the case here. There is indeed already some evidence \cite{oganesyan2007,aizenman2006} that, right at the mobility edge of the MBL transition, level statistics are likely to be Poisson. This is also consistent with our results. In Fig.\ref{fig:psbeta} we plot the exponent $\beta$, that controls the strength of level respulsion $P(s) \sim s^{\beta}, s \ll 1$, for a function of the system size $L$ around the MBL transition. A value that approaches zero in the thermodynamic limit is a signature of Poisson statistics while it is close to one for a metal. We get values of $\beta$ much closer to zero that moreover decrease with system size. This is a strong indication that at MBL transition there is no level repulsion and level statistics is Poisson.

It has been recently reported \cite{serbyn2016} that, sufficiently close to the MBL transition, the spectral correlations are scale invariant and described by semi-Poisson statistics \cite{bogomolny1999}, $P(s) \propto s e^{-As}$ with $A > 2$. Our results agree qualitatively with this picture for disorder below the transition. Level statistics in this region have indeed features typical of a disordered metal at the metal insulator-transition but there is no quantitative agreement with semi-Poisson statistics as defined in \cite{bogomolny1999}. More specifically we find a much weaker level repulsion and $A \ll 2$.    
%We do not find any signature of this statistics in our results. 
%More specifically, as was mentioned previously, we did find deviations from Poisson statistics in that region but $ A \leq  1.2$ and get very %close to the Poisson result as $h \to h_c$. Therefore we believe that at criticality the MBL transition is described by Poisson statistics. 
%For weaker disorder it was proposed in \cite{serbyn2016} that spectral correlations, still scale invariant, are described %a plasma-model \cite{kravtsov1995} with power-law interactions that predicts a decay of $P(s)$ between exponential %(Poisson) and Gaussian (WD). While we observe a qualitative similar decay of $P(s)$ in our case the spectrum is size %dependent, as expected in a metal, so we believe that it cannot be described by a scale invariant plasma model. 
We postpone a quantitative comparison with the predictions of Ref. \cite{serbyn2016} to section \ref{compa}.

 %For $h \geq 2$, still relatively far from $h_c$, the number variance is almost size independent, a signature of criticality, and very close to %Poisson statistics. This is another strong indication that criticality in the MBL problem is closely related to Poisson statistics and not to %semi-Poisson or critical statistics that describes criticality in the non-interacting limit. 

\section{Comparison between numerical results and theoretical models}\label{compa}
In this section we carry out a quantitative comparison between numerical spectral correlations and analytical results expected to describe a disordered system at the metal-insulator transition. We also explicitly compare our results with the predictions of Ref.\cite{serbyn2016}. 

We start by analysing more quantitatively the number variance close to the MBL transition.
From previous results (see Fig.~\ref{fig:variance}) it seems clear that correlations for $h \lesssim h_c$ share many of the features of an Anderson transition in the non-interacting limit: scale invariance, at least in the range of volumes that can be explored numerically, level repulsion, as in WD statistics, and linear number variance as in Poisson statistics but with a slope less than one that depends on disorder. The so called critical statistics \cite{muttalib1993,moshe1994,kravtsov1997,nishigaki1999,garcia2003} have all these features. It can be understood as a plasma model that depends on an extra parameter that controls the range of the eigenvalues interactions. Interactions for sufficiently close eigenvalues are logarithmic, as in random matrix theory, while in the opposite limit correlations are suppressed exponentially \cite{garcia2006}.   
	\begin{figure}%
		\centering
		\resizebox{0.8\textwidth}{!}{\includegraphics{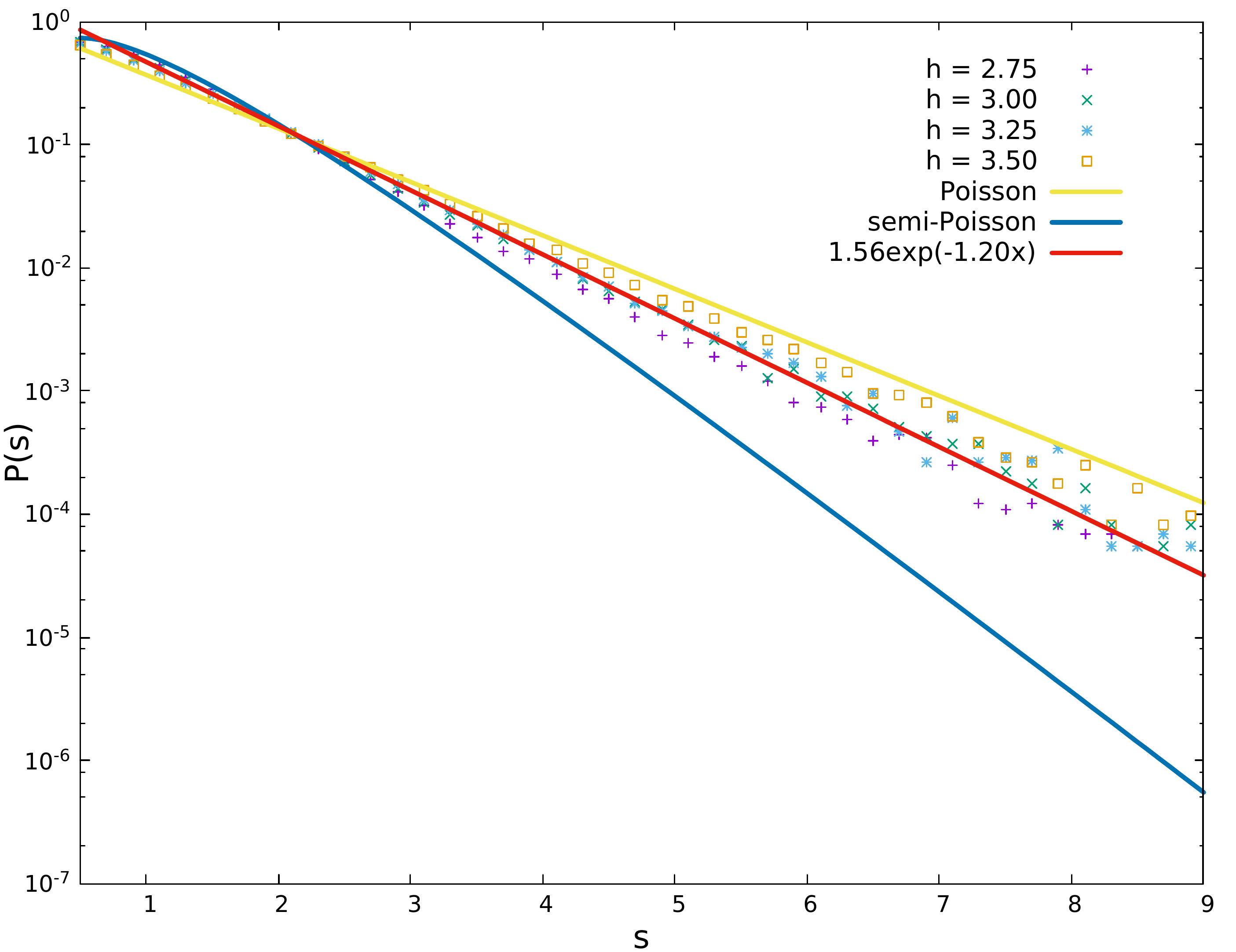}}
		\caption{Level spacing distribution $P(s)$ (\ref{eq:spacing_distribution}) close to the MBL transition $h_c \approx 3.35$. Level statistics are not close to semi-Poisson statistics for any value of $h$. The best fit for $h=3.25$ (red line), very similar to Poisson statistics, agrees with that of disordered system at the Anderson transition in the $d \gg 1$ limit.}   
		\label{fig:psemi}%
	\end{figure}%

We have fitted the numerical number variance with the prediction for critical statistics \cite{shklovskii1993,muttalib1993,moshe1994,kravtsov1997,nishigaki1999,garcia2000,garcia2002,garcia2003}. A free parameter $k$ labels the universality class which in the case of the Anderson transition depends only on the spatial dimensionality $d$ of the system. There are slightly different representations of critical statistics leading to very similar spectral correlations. Here we use the one based on the mapping onto a Calogero-Sutherland model at finite temperature \cite{garcia2003}. The final result for the number variance is: 

\begin{equation}
\label{num}
\Sigma^{2}(N)= N+2\int_{0}^{N}ds
(N-s)R_{2,c}(s).
\end{equation}

with
%\begin{equation}
%\rho(0) = \frac 1\pi \int_0^\infty dx \frac  1{1+z^{-1}e^{\frac{x^2}{2k^2N^2}}}
%\sim \sqrt{2}\frac{N\sqrt{k}}{\pi}
%\end{equation}

\begin{equation}
\label{ic}
R_{2,c}(x)= - {\bar K}^{2}
(x)-\left(\frac{d}{dx}{\bar K}(x)\right)\int_x^{\infty}{\bar K}(t)dt,
\end{equation}

\begin{equation}
\label{jk}
{\bar K}(x)=
\sqrt{k}
\int_{0}^{\infty}\frac{\cos(\pi x{\sqrt{kt}})}
{2\sqrt{t}}\frac{dt}{1+(e^{1/k}-1)^{-1}e^t}\sim \frac{\pi k}{2}\frac{\sin(\pi x)}{\sinh (\pi^2 k x/2)}
%\int_{0}^{\infty}\frac{\cos(\pi x{\sqrt{kt/2}})}
%{2\sqrt{t}}\frac{1}{{1+z e^t}}dt
\end{equation}
where strictly speaking the hyperbolic kernel only applies in the $k < 1$ limit though the phenomenology is almost identical to that of the integral kernel.
%\begin{equation}
%z^{-1}=\frac{1}{e^{1/k}-1}.
%\label{fug2}
%\end{equation}
 \begin{figure}%
 	\centering
 	\resizebox{\textwidth}{!}{
 		\resizebox{0.47\textwidth}{!}{\includegraphics{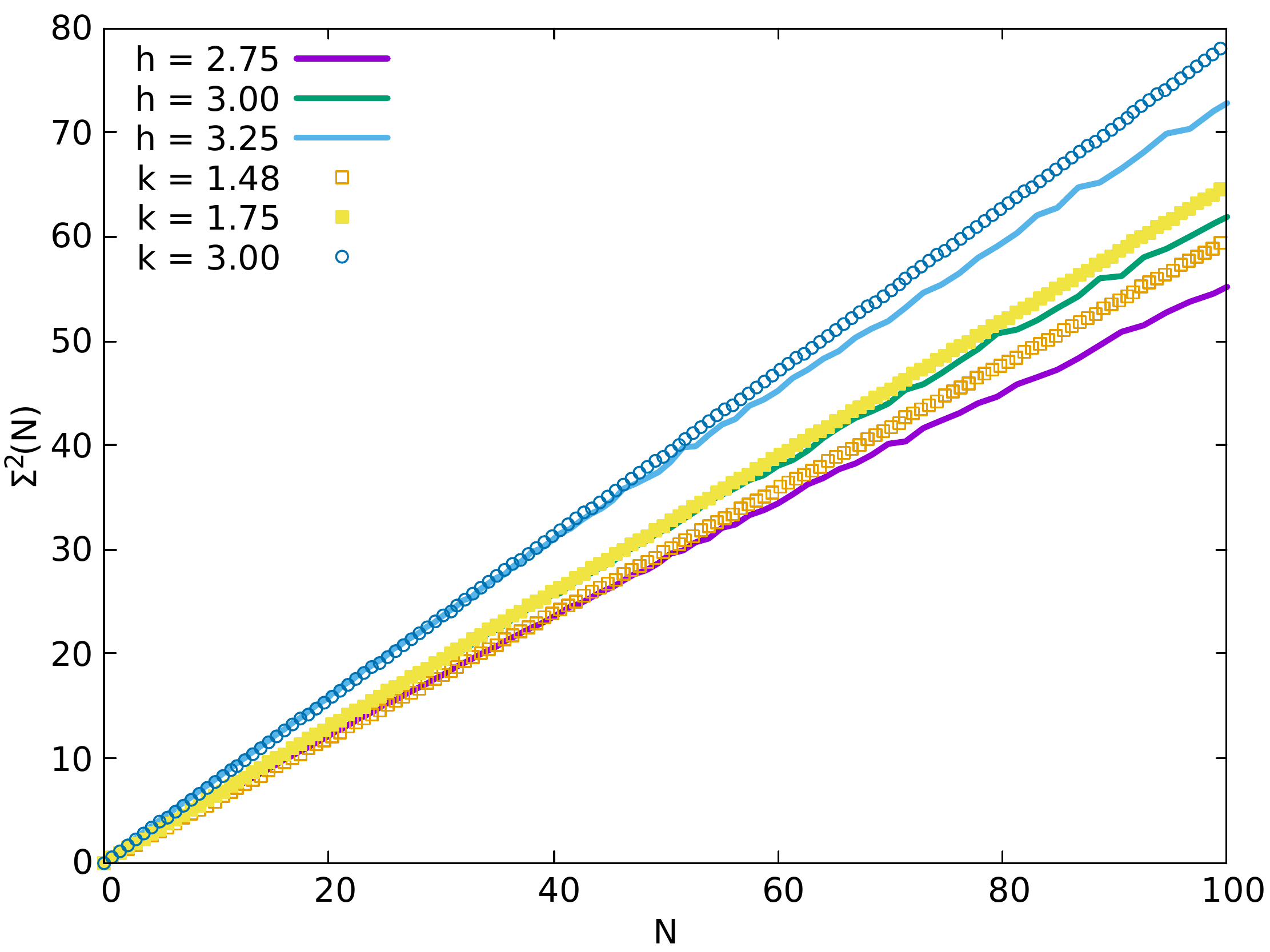}}
 		\resizebox{0.47\textwidth}{!}{\includegraphics{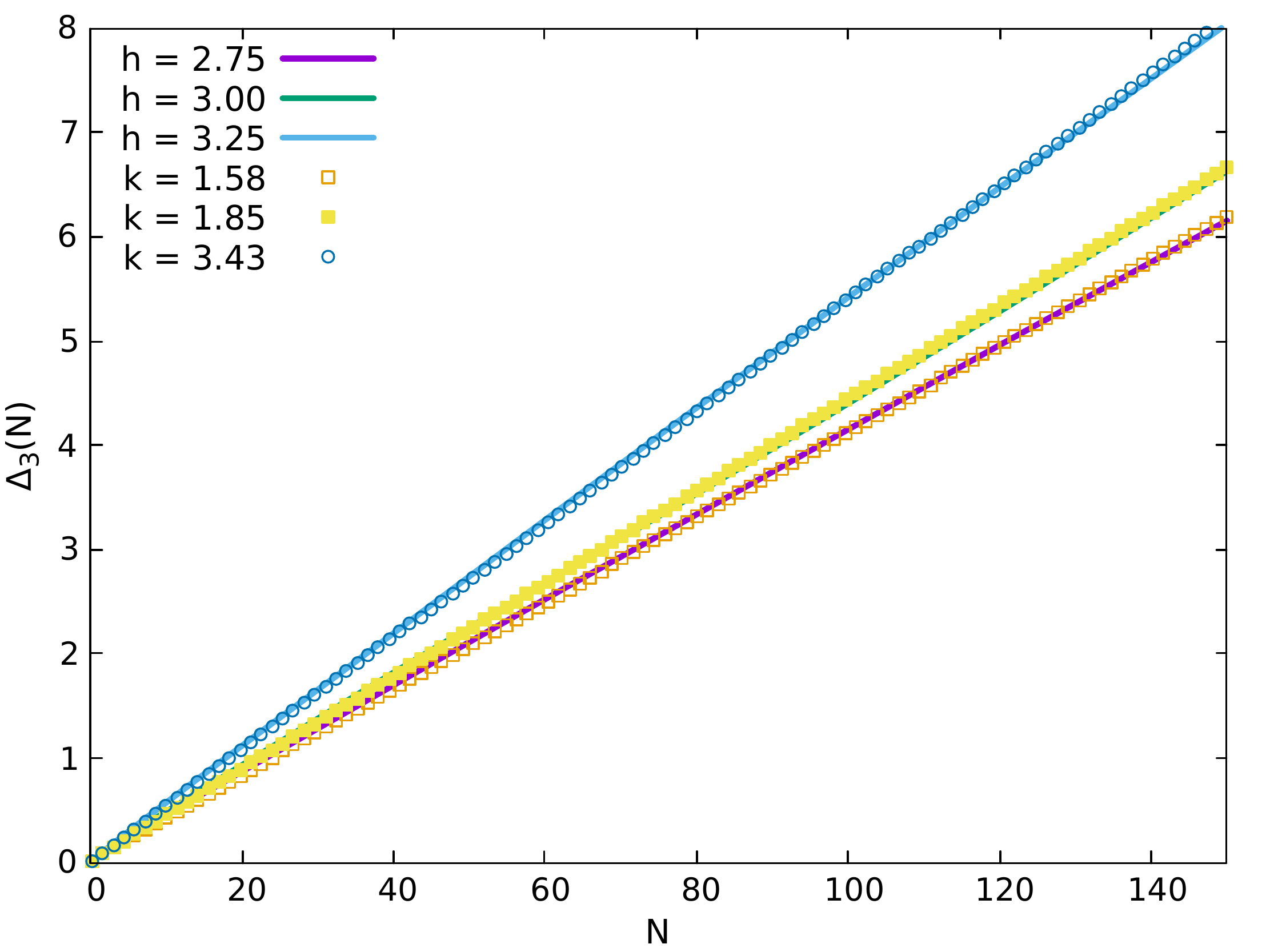}}
 		%\resizebox{0.3\textwidth}{!}{\input{variance_exponents}}
 	}
 	\caption{Left: Close to $h_c$ the number variance is well described by critical statistics. It is linear with a slope $1/2 < \chi \leq 1$, different from semi-Poisson statistics ($\chi =1/2, 1/3, \ldots$). Right: $\Delta_3(N)$ (\ref{eq:delta3}) for different disorder strength. Close to $h_c$ level statistics is well described by critical statistics $\Delta_3(N) \approx \chi N /15$. It is linear with a slope $1/2 < \chi \leq 1$, different from semi-Poisson statistics ($\chi =1/2, 1/3, \ldots$)}
 	\label{fig:variance1}%
 \end{figure}%

For all values of $k$ the number variance is asymptotically linear with a slope $ 0 < \chi(k) < 1$. 
For $h \approx h_c$ we have found good agreement, see left plot of Fig.~\ref{fig:variance1}, between the numerical results and the theoretical prediction of critical statistics. 

 Due to the limitation in the sizes that can be explored numerically it is unclear to us whether this is a genuine feature of the MBL transition or simply a size effect. It seems the latter as at $h = h_c$ level statistics are very close to Poisson $\chi = 1$ while in the non-interacting metal-insulator transition the slope is markedly smaller than one except in the $d \to \infty$ limit \cite{garcia2008}. The downward trend that we observe for large $N$ is likely a size effect 
related to the fact that we are computing the number variance for a finite number $n$ of eigenvalues so we expect \cite{brody1981} a leading correction $\sim - N^2/n$. 
 
 In order to further confirm that critical statistics describes level statistics close to the MBL transition, at least for finite lattices, we have computed $\Delta_3(N)$ statistics,
 \begin{equation}\label{eq:delta3}
 \Delta_3(N) = \frac{2}{N^4}\int_0^N(N^3-2N^2r+r^3)\Sigma^2(r)dr.
 \end{equation}
 This spectral correlator minimizes size effects since it projects out $\sim - N^2/n$ the leading size corrections \cite{brody1981}. Moreover it shares many of the number variance features: logarithmic asymptotic growth for random matrix ensembles and $\Delta_3(N) \approx \chi N/15$ for Poisson statistics with $\chi = 1$ and critical statistics $\chi < 1$. The results, depicted in Fig.~\ref{fig:variance1}, clearly show a much better agreement between the numerical spectral correlations and the predictions of critical statistics. Indeed the linear behavior, a signature of critical statistics, is still observed even for comparatively large spectral intervals. 
 
 Heuristically we could interpret these results as a progressive increase of the effective dimensionality of the system, and therefore a slope $\chi$ closer to Poisson, as the MBL transition is approached. In order words, the picture that the MBL transition is similar to the non-interacting transition in $d \to \infty$ could be generalized to the metallic phase close to the transition where the dynamics, at least for small sizes, is similar to that of the critical region in a disordered non-interacting metal at large but finite dimension. That is also consistent with the fact \cite{garcia2008} that the slope of the number variance at the Anderson transition $\chi \sim 1 - 2/d$ tends to  Poisson statistics in the $d \to \infty$ limit.
  Indeed intriguing similarities between two similar problems,  the MBL problem and the problem of a single particle in a Cayley tree that resembles a regular lattice in the $d \to  \infty$ limit,  have already been suggested in the literature \cite{basko2006}.  Moreover rigorous results in the mathematical \cite{aizenman2007} literature suggest that, provided that the MBL problem can be approximately mapped to a Bethe lattice, Poisson statistics will describe level statistics at the MBL transition. Numerical simulations  in small lattices \cite{serbyn2016,oganesyan2007} have confirmed that level statistics is close to Poisson at the MBL transitio. Therefore it is reasonable to expect that, for a slightly below disorder, the MBL problem must be related to the properties of a high, but finite, dimensional disordered conductor.  
  \begin{figure}%
  	\centering
  	\resizebox{\textwidth}{!}{
  		\resizebox{0.8\textwidth}{!}{\includegraphics{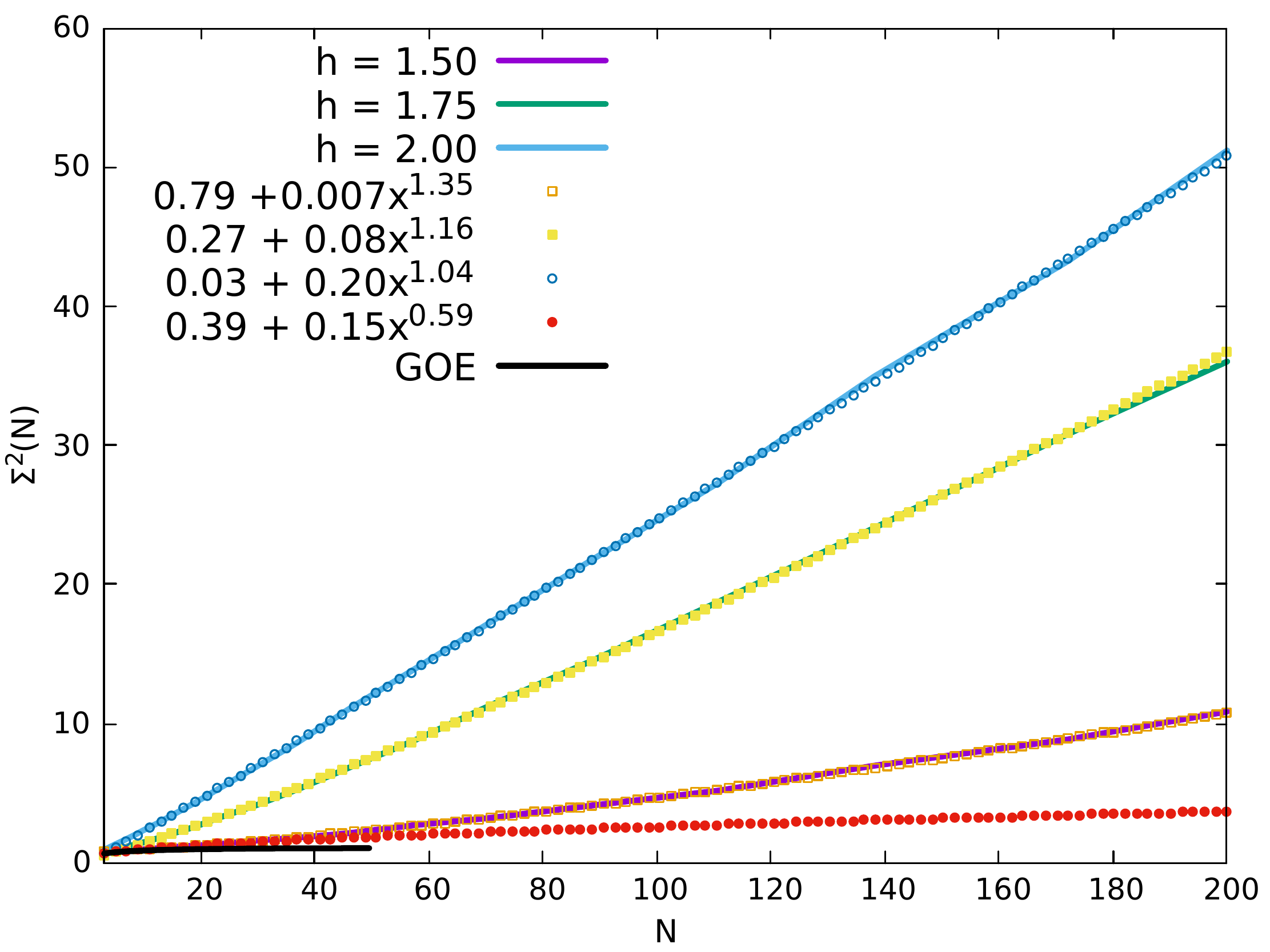}}
  		%\resizebox{0.9\textwidth}{!}{\includegraphics{nvcri.pdf}}
  		%\resizebox{0.3\textwidth}{!}{\input{variance_exponents}}
  	}
  	\caption{$\Sigma^2(N)$ (\ref{eq:nv}) for different intermediate disorder strength well in the metallic phase. The growth of the number variance is faster than linear in agreement with the existence of a Thouless energy in the system. In all cases the fitting interval is $[20,200]$. The best fitting is not very sensitive on the fitting interval provided that the lower limit is $N \gg 1$ and the upper limit $100 < N < 300$. If the interval is too small, $[3,20]$ (red circles), the best fitting is very sensitive to the fitting interval. Moreover the predicted slower than linear growth (red dots) seems to be an artefact of the fitting procedure. The numerical results (purple line) and the modified fitting (orange squares) clearly points to a much faster growth. We believe that this is the reason for the disagreement with the results of Ref.\cite{serbyn2016}. For comparison we also show results for the Gaussian Orthognal Ensemble (GOE) where spectral rigidity (\ref{eq:nvwd}) is observed at all scales.}
  	\label{fig:variance2}%
  \end{figure}%

 We now compare explicitly the number variance for intermediate disorder $1.5 < h < 2.5$ with the prediction of Ref. \cite{serbyn2016}: a plasma model with power-law interactions \cite{kravtsov1995} leading to a growth of the number variance slower than linear. The results, depicted in Fig.~\ref{fig:variance2}, clearly show that for a sufficiently large $N$ the growth of the numerical number variance is faster than linear, a signature of the Thouless energy of the system.  Moreover the numerical level statistics are not scale invariant so it cannot in principle be described by a scale invariant plasma model. It seems more likely the numerical spectral correlations are similar instead to that of a plasma-model \cite{jalabert1993} with short-range logarithmic correlations and long-range, size-dependent, power-law correlations.  We believe that the discrepancy with our results has to do with the fact that the fitting carried out in \cite{serbyn2016} only involved small $N  \leq 20$. In order to observe the faster than linear growth it is necessary to explore a substantially larger interval. 
  
  Finally we comment on the recent claim \cite{serbyn2016} that very close to the MBL transition, but still on the metallic side, spectral correlations are similar to that of a  disordered system at the Anderson transition and are well described by semi-Poisson statistics \cite{bogomolny1999}. As was mentioned previously, semi-Poisson statistics \cite{bogomolny1999} describes the spectral correlations of an eigenvalue plasma model where interactions are restricted to nearest neighbours only. The number variance and level spacing distribution have a particularly simple form,
  \begin{eqnarray}
  	\Sigma^2(N)= N/2 + \frac{1 - e^{-4N}}{8}, ~~ P(s)=4se^{-2s}. 
  \end{eqnarray}
 This intermediate statistics describes well the spectral correlations of pseudo integrable billiards and some qualitative features of the Anderson transition. We agree with \cite{serbyn2016} that level statistics close to the MBL transition is similar to that of system at the metal-insulator transition but we do not observe any quantitative signature of semi-Poisson statistics in the numerical spectral correlations. The number variance, depicted in Fig.~\ref{fig:variance1} for $L=18$ for different $h \approx h_c$, is linear but quite close to Poisson statistics $\Sigma^2(N)= N$. The slope increases with $h$ and is above the semi-Poisson statistics prediction $\chi \leq 1/2$ for $h > 2.6$. 
  Therefore close to $h_c$ level statistics are never well described by semi-Poisson statistics. 
  Results for $P(s)$, see Fig.~\ref{fig:psemi}, confirm that level statistics are much closer to Poisson, and better described by critical statistics, than to semi-Poisson statistics in the region close to the MBL transition. Taking into account that no explicit comparison with semi-Poisson statistics \cite{bogomolny1999} was carried out in \cite{serbyn2016}, we believe that the source of disagreement are not the numerical results of \cite{serbyn2016} close to the MBL transition but on the terminology used. It seems that semi-Poisson statistics in \cite{serbyn2016} is not used in the strict sense of the plasma model of \cite{bogomolny1999} but rather in a broader sense to refer to level statistics at the Anderson transition. 

\section{Conclusions}

We have investigated the interplay of interactions and disorder by level statistics in a one dimensional XXZ spin chain in a random field. In agreement with previous works, we have found a transition from a metallic phase to a many-body localized phase at a finite disorder $h \approx 3.4$. The critical exponent $\nu \approx 1$ that controls the MBL transition, here obtained by level statistics, is not far form that of a non-interacting particle in a Cayley tree that mimics a conventional lattice in the $d \to \infty$ limit.  
Deep in the metallic phase we have found for the first time a clear evidence of the existence of the Thouless energy in this system.  As expected in a metal, for eigenvalues separated less than the Thouless energy, level statistics are well described by WD statistics. For larger separations, the number variance shows a growth faster than linear consistent with quantum dynamics governed by a process of anomalous diffusion. Additional research would be necessary to determine exactly the precise relation between this anomalous Thouless energy and the subdiffusive phase that occurs on the metallic but not far from the MBL transtion. As the system approaches the MBL transition, spectrum becomes approximately scale invariant and no Thouless energy is observed. The number variance and the level spacing distribution gradually becomes closer to Poisson statistics. This result is consistent with a metal-insulator transition in the limit of infinite spatial dimensions. Slightly below the MBL transition level statistics are well described by critical statistics. Spectral correlations in this limit are similar to that of a high-dimensional, non-interacting disordered system at the Anderson transition. It is likely that eigenstates in this regime are sparse multifractal with typical exponent that are sensitive to disorder. 

\acknowledgments
A. M. G. acknowledges support
from EPSRC, Grant No. EP/I004637/1.
\appendix

\section{Unfolding}
\label{ap:unfolding}

\begin{figure}%
	\centering
	\begin{subfigure}[a]{0.8\textwidth}
		\resizebox{\columnwidth}{!}{\includegraphics{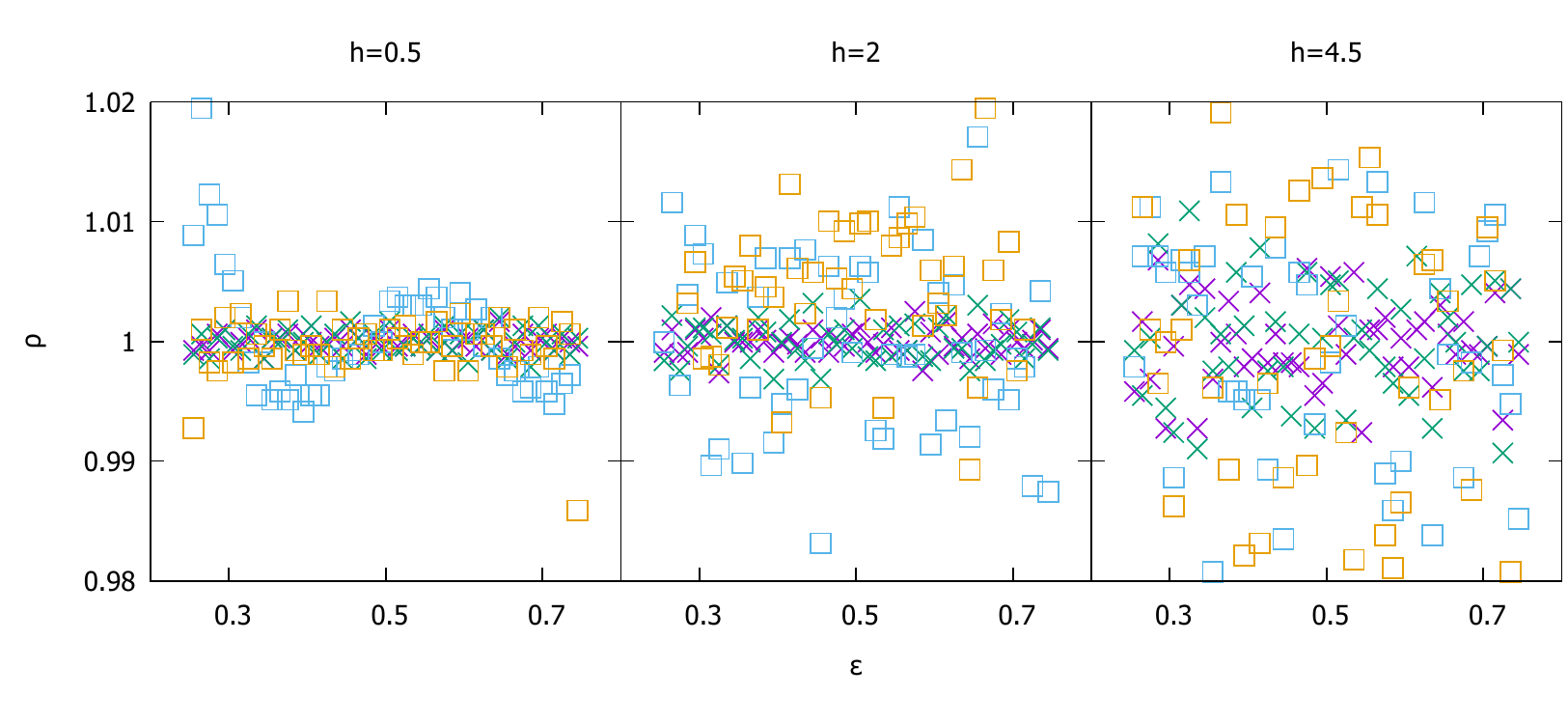}}
	\end{subfigure}
	\begin{subfigure}[b]{0.8\textwidth}
		\resizebox{\columnwidth}{!}{\includegraphics{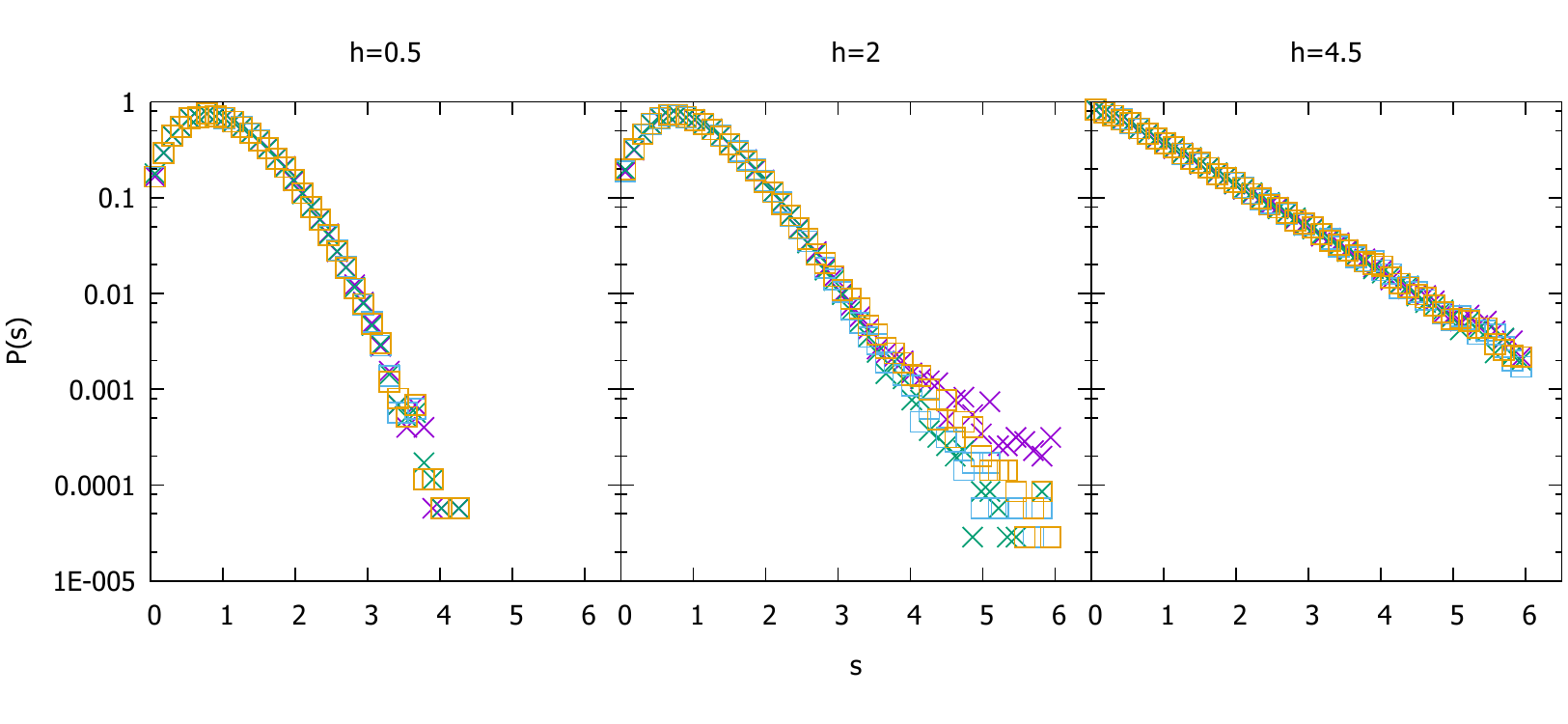}}
	\end{subfigure}
	\begin{subfigure}[c]{0.8\textwidth}
		\resizebox{\columnwidth}{!}{\includegraphics{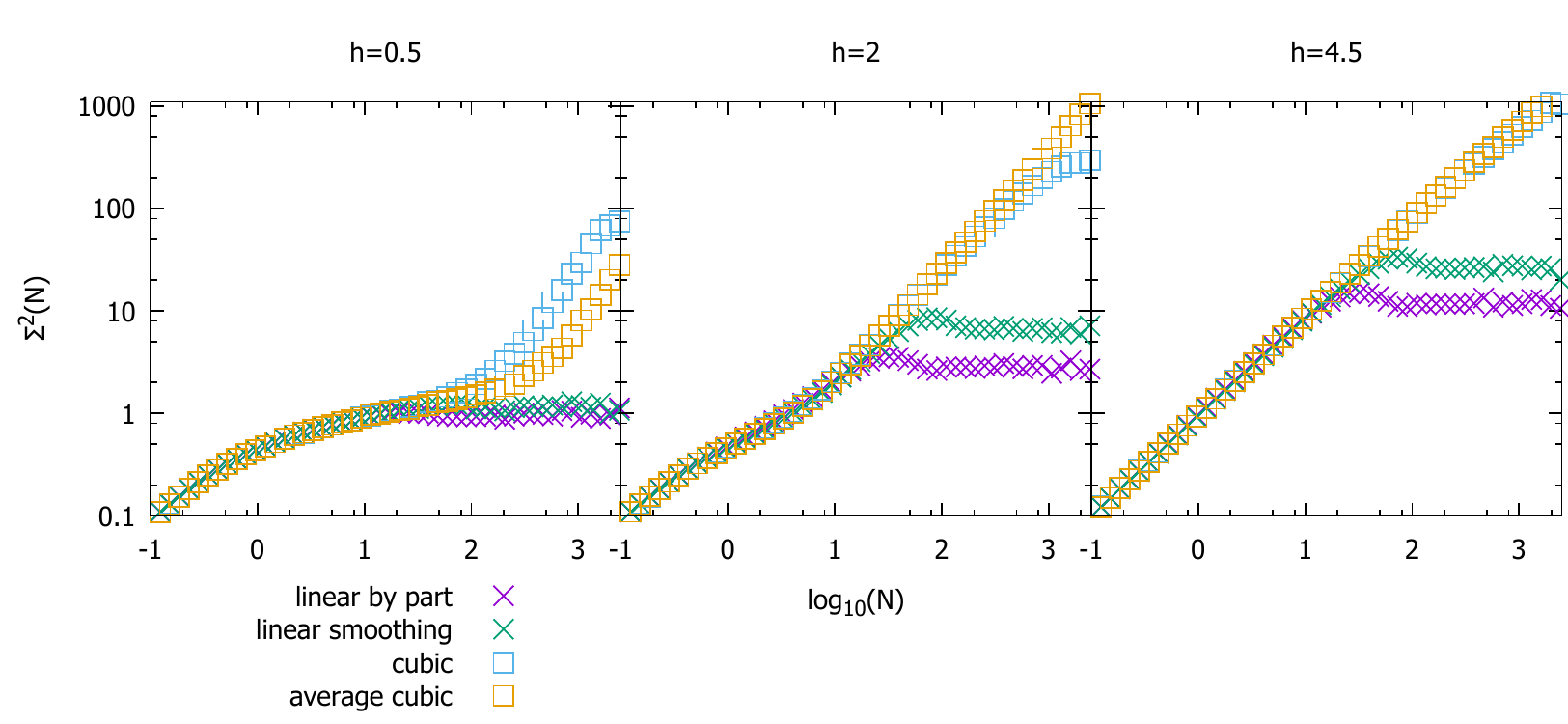}}
	\end{subfigure}
	\caption{Comparison between the impact of local (crosses) and global (squares) unfoldings (definitions in appendix~\ref{ap:unfolding}) on the level statistics for $L=18$. The density of states $\rho$ shows local methods are more accurate, but they destroy correlations in the number variance $\Sigma^2(N)$. Local and global methods agree on the spacing distribution $P(s)$ and on short-range correlations.}%
	\label{fig:comparison_unfolding}%
\end{figure}

This appendix presents in more details the unfolding theory and defines the unfolding methods which have been considered in section~\ref{sec:unfolding} and compared in Fig.~\ref{fig:comparison_unfolding}.\\

To use level statistics, it is necessary to unfold the part of the spectrum considered. Indeed, to compare statistics from different parts of the spectrum, their densities of state must be equal. Unfolding consists in stretching the spectrum to normalise the density of state to unity. This is done by a fit of the staircase function $\eta$ \cite{guhr1998}, defined for increasingly sorted eigenvalues $E_1,...,E_N$ as:
\begin{equation}
\begin{split}
\eta(E) &= \int_{-\infty}^E S(E')dE' = \sum_{n=1}^N \Theta(E-E_n) \\ 
S(E) &= \sum_{n=1}^N \delta(E-E_n) \\
\end{split}
\end{equation}
The staircase function is decomposed in a smooth part $\bar{\eta}(E)$ and fluctuations $\delta\eta(E)$:
\begin{equation}
\eta(E) = \bar{\eta}(E) + \delta\eta(E)
\end{equation}
The slope of the smooth part gives the local density of state.
Unfolding correspond to mapping the eigenvalues onto the smooth part:
\begin{equation}
E_n \rightarrow \epsilon_n =  \bar{\eta}(E_n)
\end{equation}
This change of variable has transformed the staircase function into:
\begin{equation}
\hat{\eta}(\epsilon) = \epsilon + \delta\hat{\eta}(\epsilon)
\end{equation}
Where it is seen that the density of states is unity over the whole spectrum.\\

The difficulty lies in the definition of the smooth part. There are as many ways of unfolding as there are fitting or smoothing methods. However, the unfolding methods are divided between the local and the global methods. A local unfolding calculates $\bar{\eta}(E)$ from the values of $\eta$ in a range $\left[E+\Delta E, E-\Delta E\right]$ thus deleting any correlations between levels separated by more than $2\Delta E$. As a result, the correlation function and any observable related -- such as the number variance -- are irrelevant for high energies. On the contrary, a global unfolding using all values of $\eta$ preserves these correlations, but it is more difficult to check its validity.\\

Four different methods have been tested and compared. The simplest uses a global polynomial regression with degree 3. Higher degrees can lead to over-fitting. Degree 1 (linear regression) gives the average spacing from the invert of the resulting slope. The other global unfolding implemented is the most complex. It assumes $\bar{\eta}(E)$ is defined as the average staircase function (averaged over all disorder realisations). As a consequence, a polynomial regression of degree 3 is used to fit an estimate of the average staircase function, result of the average over all disorder realisation calculated so far. The complexity comes from this averaging. Because eigenvalues are sorted in the staircase function, its shape varies a lot with different disorder realisations, which makes it difficult to average correctly with a small number of realisations (a few dozen in our case for $L=18$). It can be seen experimentally that these staircase functions can be rescaled into a single shape with a change of variable $E \rightarrow k(E-E^0)$ where $k$ depends on the disorder realisation. After this transformation the averaging is more efficient.\\

The first local unfolding method is a simple non-continuous linear by part fitting. The spectrum is divided in blocks of 100 consecutive eigenvalues, and each block is fitted independently with a linear regression. The second local method is called linear smoothing. For each eigenvalue, the block of $100$ levels around the eigenvalue of interest is fitted with a linear regression. Only the eigenvalue of interest is mapped with this fit. As a result there is a different fit for mapping each level.\\
The comparison of these four methods (Fig.~\ref{fig:comparison_unfolding}) shows that local methods average the density of state $\rho$ to unity in a finer way than global methods. Moreover the simple cubic method gives a ``bad'' density of state, proof that it does not unfold correctly. Nevertheless, global unfoldings keep the correlations at any range, and seem to agree with local unfolding on short-range correlations and spacing distribution. Because long-range correlations are needed in this work, we assumed that the global averaging unfolding method is correct.

\section{Scaling analysis}
\label{ap:scaling_analysis}

The critical disorder $h_c$ and critical exponent $\nu = 1 / \mu$ have been estimated using a scaling analysis on the adjacent gap ratio. The idea is to extrapolate the ratio from different system sizes to an infinite volume. For $L=12, 14, 16, 18$ and $h=2, 2.25, 2.5, ..., 4$ a global value $R_L(h)$ has been calculated from the average adjacent gap ratio over the 4\% centre of the spectrum. Each $R_L$ have then been rescaled with the change of variable $h \rightarrow (h-h_c)L^{\mu}$, so that the different $R_L$ form a single curb\cite{luitz2015}. In practice, $h_c$ and $\mu$ are found by minimizing the cost function:
\begin{equation}
% S(h_c, \mu) = \sum_{i=1}^N \frac{Var_L(R_L((h_i-h_c)L^\mu))}{h_{\rm max} - h_{\rm min}}
S(h_c, \mu) = \frac{1}{h_{\rm max} - h_{\rm min}} \int_{h_{\rm min}}^{h_{\rm max}} Var_L(R_L((h-h_c)L^\mu)) dh
\label{eq:sc}
\end{equation}
where $Var_L$ is the variance over the different values of $L$. $R_L$ are interpolated using cubic splines. In order to estimate the error in $h_c$ and $\nu$ we carry out the fitting in the random interval where $h_{\rm min}$ is extracted from a box distribution between $2$ and $3$ and $h_{\rm max}$ between $3.75$ and $4$. The starting values of $h_c$ and $\mu$ are taken from a box distribution $[2.75, 3.75]$ and $[0.5, 2]$ respectively. The error in $\nu$ and $h_c$ results from the standard deviation after averaging over the above different fitting intervals.
\bibliography{library}

\end{document}